\documentclass[conference]{IEEEtran}
\IEEEoverridecommandlockouts
\usepackage{cite}
\usepackage{amsmath,amssymb,amsfonts}
\usepackage{algorithmic}
\usepackage{graphicx}
\usepackage{textcomp}
\usepackage{xcolor}

\usepackage{cleveref}
\usepackage[color=lightgray]{todonotes}
\usepackage[ruled, vlined, linesnumbered]{algorithm2e}
\usepackage{subcaption}
\usepackage{multirow}
\usepackage{bm}
\usepackage{soul}
\usepackage{bbold}
\usepackage{mathtools}
\usepackage[utf8]{inputenc}
\usepackage{booktabs}
\usepackage{listings}
\usepackage{tabularx}
\usepackage{makecell}
\usepackage{pifont}
\usepackage{siunitx}
\crefname{lstlisting}{algorithm}{algorithms}
\Crefname{lstlisting}{Algorithm}{Algorithms}

\newcommand{\markblue}[1]{\textcolor{black}{#1}}
\newcommand{\ours}{{MetaRec}\xspace}

\def\BibTeX{{\rm B\kern-.05em{\sc i\kern-.025em b}\kern-.08em
    T\kern-.1667em\lower.7ex\hbox{E}\kern-.125emX}}
\begin{document}

\title{Transferable Sequential Recommendation via Vector Quantized Meta Learning}

\author{
\IEEEauthorblockN{
Zhenrui Yue\textsuperscript{1}, Huimin Zeng\textsuperscript{1}, Yang Zhang\textsuperscript{1}, Julian McAuley\textsuperscript{2}, Dong Wang\textsuperscript{1}
}
\IEEEauthorblockA{
\textsuperscript{1}University of Illinois Urbana-Champaign, Champaign, IL, USA \\
\textsuperscript{2}UC San Diego, San Diego, USA \\
\{zhenrui3, huiminz3, yzhangnd, dwang24\}@illinois.edu, jmcauley@ucsd.edu}
}

\maketitle

\begin{abstract}
\markblue{
While sequential recommendation achieves significant progress on capturing user-item transition patterns, transferring such large-scale recommender systems remains challenging due to the disjoint user and item groups across domains. In this paper, we propose a vector quantized \ul{meta} learning for transferable sequential \ul{rec}ommenders (\ours). Without requiring additional modalities or shared information across domains, our approach leverages user-item interactions from multiple source domains to improve the target domain performance. To solve the input heterogeneity issue, we adopt vector quantization that maps item embeddings from heterogeneous input spaces to a shared feature space. Moreover, our meta transfer paradigm exploits limited target data to guide the transfer of source domain knowledge to the target domain (i.e., learn to transfer). In addition, \ours adaptively transfers from multiple source tasks by rescaling meta gradients based on the source-target domain similarity, enabling selective learning to improve recommendation performance. To validate the effectiveness of our approach, we perform extensive experiments on benchmark datasets, where \ours consistently outperforms baseline methods by a considerable margin.
}
\end{abstract}


\section{Introduction}

% sequential recommenders and transfer learning
Thanks to recent advances in language modeling, sequential recommendation has experienced significant improvements in capturing user-item transition patterns~\cite{kang2018self, sun2019bert4rec, li2021share, zhou2022filter, yue2023llamarec, zeng2024federated}. While sequential recommendation outperforms traditional methods, a common challenge is that well-trained models cannot be reused for an unseen domain. As such, transferable recommenders are proposed for quick adaptation to a different target domain~\cite{singh2008relational, tang2012cross}. One popular approach is to leverage shared information across domains (e.g., shared items) to enhance adaptation performance~\cite{zhang2018cross, yuan2021one, zhu2022personalized}. Another stream of cross-domain recommendation aims at learning domain-invariant features~\cite{zhu2018deep, wang2019recsys, li2022recguru}. However, the mentioned approaches assume (partially) overlapping user~/~item groups or require explicit correspondences. Therefore, they are inapplicable upon large source-target domain discrepancy~\cite{ding2021zero}. Recently, transfer learning methods are proposed by utilizing auxiliary information, where descriptive input (e.g., item title) is encoded as features~\cite{wang2022transrec, hou2022towards, li2023text}. Yet current approaches may cause performance drops due to the over-emphasis of domain-specific features~\cite{hou2022learning}. Additionally, such methods require additional input, rendering them less effective in text-scarce or sensitive domains.

% our setting and challenges
To generalize transferable sequential recommenders to universal model architectures and recommendation scenarios, we consider an \textit{ID-only}, \textit{non-overlapping} and \textit{multi-source} transfer learning setting, where items are solely represented with IDs and user interaction histories are sequences of item IDs in chronological order. Moreover, we assume zero overlapping of shared information across domains, that is, the involved domains only comprise of \textit{mutually exclusive user and item groups}. Consequently, our approach enables the transfer of knowledge from arbitrary source domains to a different target domain in spite of the input heterogeneity, which significantly extends the applicability of cross-domain recommendation. The primary challenge of this setting is twofold: (1)~the disjoint input spaces and item-level differences can lead to alignment difficulties across different domains; and (2)~user behavior patterns in source domains may differ from those in the target domain, potentially causing negative transfer (e.g., performance drop) upon large domain discrepancy.

% our method
To this end, we propose vector quantized \ul{meta} learning for universally transferable sequential \ul{rec}ommenders (\ours). \ours can accommodate arbitrary recommender architecture and consists of: (1)~vector quantization (VQ) and (2)~meta transfer. VQ solves the input heterogeneity problem by mapping the original item embeddings to a shared feature space. Instead of introducing additional parameters, we apply weights from the target domain embedding table as codebook in VQ. Then, the output vectors in the aligned feature space are used as item features to predict the next interaction. Despite quantizing the item representations, transition patterns from the source domains may be of different similarity to the target domain. Therefore, we additionally design meta transfer that adaptively learns to transfer knowledge from data-intensive source domains to the data-scarce target domain. Specifically, we update the parameters with sampled source domain data (i.e., source tasks), followed by deriving meta gradients using sampled target domain examples. Based on source-target gradient similarity, we rescale the meta gradients to optimize the learning from different source tasks. As such, \ours learns domain-invariant features from source optimization paths with the objective of improving the target domain performance. We summarize our contributions below:
\begin{enumerate}
\item To the best of our knowledge, we are the first to propose a solution for cross-domain sequential recommendation based on an \emph{ID-only} setting with \textit{disjoint} item groups.
\item The proposed vector quantization maps item embeddings across domains to a well-aligned feature space. Moreover, our meta transfer `learns to transfer' from multiple sources for improved target domain performance.
\item We demonstrate the effectiveness of our \ours with extensive experiments over multiple source-target dataset selections, where the proposed \ours consistently outperforms baseline methods with considerable improvements in recommendation performance.
\end{enumerate}
\section{Related Work}

\subsection{Transferable Recommender Systems}
Transferable recommender systems are proposed to improve performance in data-scarce settings~\cite{cantador2015cross}. There exist several approaches to address transfer learning, with the majority of existing work relying on the assumption of shared user or item groups~\cite{zhu2022personalized, chen2023toward}. Another possible approach is to leverage generalized representations for improved knowledge transfer~\cite{li2021dual, li2022recguru}. Recently, auxiliary features are proposed to improve cross-domain recommendation performance, where additional modalities (e.g., text) or auxiliary information (e.g., item category) are adopted to generate item features based on the associated descriptive input~\cite{li2023text, wang2024train}. However, existing approaches either require shared information across domains or use additional auxiliary information for recommendation. Therefore, we propose a generalized transfer learning framework that solely relies on item IDs to extend the applicability of transferable sequential recommenders.

\subsection{Vector Quantization}
Vector quantization (VQ) refers to mapping high-dimensional vectors to discrete codes using prototype vectors, which are often learnt in an unsupervised fashion~\cite{kohonen1995learning}. Recently, VQ is known for being used in generative models for discrete representations in the latent space~\cite{van2017neural}. Vector quantization is also applied to learn compact features or semantic IDs for improved efficiency and performance in recommenders~\cite{van2019pq, rajput2023recommender}. For cross-domain recommendation, VQ-Rec proposes to leverage VQ that generates discrete codes upon textual features~\cite{hou2022learning}. Yet previous VQ methods focus on improving recommendation efficiency or domain-invariant semantic IDs, the potential of vector quantization in learning domain-invariant item features remains unexplored for sequential recommender systems.

\subsection{Meta Learning}
Meta learning (i.e., learning to learn) demonstrates superior performance in few-shot learning, where limited training examples are provided for the desired task~\cite{li2016learning}. A common meta learning approach is to construct a meta-learner that guides the optimization of the learner's parameters~\cite{andrychowicz2016learning, ha2016hypernetworks}. For example, model-agnostic meta learning (MAML) uses second-order optimization to learn initial parameters that quickly adapt to a new task~\cite{finn2017model}. In recommender systems, meta learning methods have also been applied to improve data-scarce recommendation or enhance recommendation fairness~\cite{lee2019melu, qin2023meta}. Unlike previous works, we consider cross-domain setting and propose meta learning-based transferable recommendation, which `learns to transfer' from multiple sources to optimize the target domain recommendation.
\section{Methodology}

\subsection{Framework}

% setup
Our transfer learning framework is based on sequential recommendation, in which user interaction history $\bm{x}$ is used as input. Specifically, $\bm{x}$ is a sequence of user interactions $[x_1, x_2, \ldots, x_T]$ with length $T$ in chronological order, where the items are represented with unique IDs in the item space $\mathcal{I}$ (i.e., $x_i \in \mathcal{I}, i = 1, 2, \ldots, T$). The output of the recommender is the prediction scores $\hat{y} \in \mathbb{R}^{|\mathcal{I}|}$ over input space $\mathcal{I}$, whereas the ground truth item is denoted with $y \in \mathcal{I}$. We consider the multi-source transfer learning setting:
\begin{itemize}
\item \emph{Source Domain}: Let $\{ \mathcal{D}^{s}_{i} \}^{M}_{i=1}$ be the set of $M$ source domains ($M > 1$), each source domain $\mathcal{D}^{s}_{i}$ is defined by its item space $\mathcal{I}^{s}_{i}$, user group $\mathcal{U}^{s}_{i}$ and dataset $\mathcal{X}^{s}_{i}$, in which $|\mathcal{U}^{s}_{i}|$ data examples are provided (i.e., $\mathcal{X}^{s}_{i} = \{ (\bm{x}^{s}_{ij}, y^{s}_{ij}) \}^{|\mathcal{U}^{s}_{i}|}_{j=1}$). All source domain datasets are available to facilitate the transfer learning process.
\item \emph{Target Domain}: Similar to the source domain, we define the target domain $\mathcal{D}^{t}$ with item space $\mathcal{I}^{t}$, user group $\mathcal{U}^{t}$ and dataset $\mathcal{X}^{t}$. Target data $\mathcal{X}^{t} = \{ (\bm{x}^{t}_{j}, y^{t}_{j}) \}^{N^{t}}_{j=1}$ is also provided (smaller than source datasets). To make our setting universally applicable, we additionally assume non-overlapping user and item groups across all domains, i.e., $\mathcal{I}^{s}_{i} \cap \mathcal{I}^{t} = \emptyset, \mathcal{U}^{s}_{i} \cap \mathcal{U}^{t} = \emptyset,$ with $i = 1, 2, \ldots, M$.
\end{itemize}
We denote the recommender model with $\bm{f}$, which is parameterized by $\bm{\theta}$ (i.e., $\hat{y} = \bm{f}(\bm{\theta}; \bm{x})$). The model $\bm{f}$ comprises a embedding table (denoted with $\bm{f}_{e}$) and an encoding model $\bm{f}_{m}$, with $\bm{f} = \bm{f}_{m} \circ \bm{f}_{e}$. Ideally, the highest ranked item in $\hat{y}$ matches the ground truth $y$ (i.e., $y = \arg\max \hat{y}$). The objective of our framework is to optimize the target domain performance. In other words, we seek to minimize the expectation of recommendation loss $\mathcal{L}$ w.r.t. parameters $\bm{\theta}$ over $\mathcal{X}^{t}$:
\begin{equation}
    \label{eq:transfer_objective}
    \min_{\substack{\bm{\theta}}} \mathbb{E}_{(\bm{x}^{t}, y^{t}) \sim \mathcal{X}^{t}} [\mathcal{L}(\bm{f}(\bm{\theta}; \bm{x}^{t}), y^{t})].
\end{equation}

\subsection{Vector Quantization}

% codebook
\subsubsection{Mapping Quantized Representations}
\begin{figure}[t]
    \centering
    \includegraphics[trim=5.9cm 4.5cm 7.05cm 4.8cm, clip, width=0.9\linewidth]{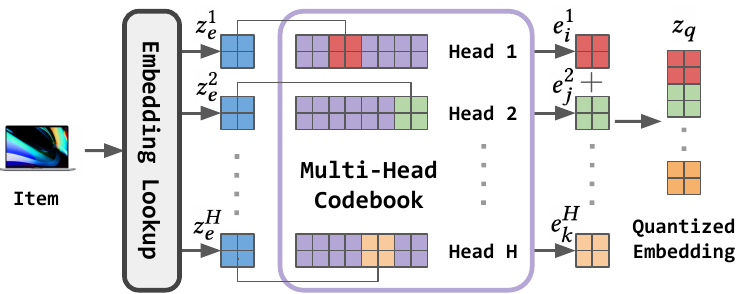}
    \caption{Scheme of our vector quantization module. The item embeddings $z_{e}$ is split into $H$ heads and projected separately.} 
    \label{fig:vq}
    \vspace{-10pt}
\end{figure}

Our VQ module consists of a multi-head codebook $\bm{e} \in \mathbb{R}^{H \times K \times D}$, where $H, K, D$ represent the head number, codebook size and hidden dimension. To avoid introducing additional parameters and overfitting to the limited target data, we reshape the target domain embedding table to obtain the codebook $\bm{e}$. Consequently, $K$ is the size of target domain items $|\mathcal{I}^{t}|$ and $H \times D$ is equal to the hidden dimension of the model $\bm{f}$. Here, we denote the $j$-th embedding vector of the $i$-th head as $\bm{e}^{i}_{j} \in \mathcal{R}^{D}$, with $i \in 1, 2, \ldots, H$ and $j \in 1, 2, \ldots, K$. For simplicity, the embedding of item $x$ is referred as $z_{e}$ in the following discussion (i.e., $z_{e} = \bm{f}_{e}(x)$). We split the hidden dimension of $z_{e}$ into $H$ equal dimensions (i.e., $z_{e} = [z^{1}_{e}; z^{2}_{e}; \ldots; z^{H}_{e}]$), as we find it beneficial to split item representations into subspaces and project them separately (See \Cref{fig:vq}). Additionally, multi-head VQ can increase the total number of vector combinations from $K$ to $K^{H}$, significantly increasing the output space size of the quantized embeddings. In particular, we denote the mapping for item $x$ or its embedding $z_e$ with $\bm{f}_{q}$: 
\begin{equation}
  \begin{aligned}
    z_{q} &= \bm{f}_{q}(z_{e}) = \bm{f}_{q}(\bm{f}_{e}(x)) = [e^{1}_{i}; \, e^{2}_{j}; \, \ldots; \, e^{H}_{k}], \, \mathrm{where} \\
    i &= \arg\max_{l} \mathrm{Sim} (z^{1}_{e}, e^{1}_{l}), j = \arg\max_{l} \mathrm{Sim} (z^{2}_{e}, e^{2}_{l}), \ldots,
  \end{aligned}
  \label{eq:vq_function}
\end{equation}
in which $z_{q}$ is the quantized embedding and Sim denotes cosine similarity (Sim$(a, b) = \frac{a \cdot b}{\| a \| \| b \|}$). In $\bm{f}_{q}$, we select cosine similarity to compute the nearest elements of $z_{e}$ from the codebook vectors head-wise. Next, the nearest elements (i.e., codebook vectors with highest similarity) in each of the $H$ heads are concatenated to obtain the quantized embedding, and the indices $i, j, \ldots$ in \Cref{eq:vq_function} are the semantic codes.

% learning of vq
\subsubsection{Learning Quantized Representations}
Since the quantized embedding $z_{q}$ is used as input to encoder model $\bm{f}_{m}$, $\nabla_{z_{q}} \mathcal{L}$ w.r.t. $z_{q}$ can be obtained in backpropagation. Yet $\bm{f}_q$ is a discrete mapping function, thus $z_{e}$ can not be directly learnt with gradient descend. As a solution, we instead pass the gradients $\nabla_{z_{q}} \mathcal{L}$ to $z_{e}$ to optimize the item embeddings~\cite{van2017neural}. Training $z_{q}$ in $\bm{e}$ corresponds to the learning of both domain-invariant centroid vectors and target domain embeddings (Note $\bm{e}$ shares weights with the target domain embeddings). For this purpose, we update $z_{q}$ by maximizing the cosine similarity between matched pairs of $z_{q}$ and $z_{e}$. Notice that the closed-form solution for $\max_{z_{q}} \mathrm{Sim} (z_{q}, z_{e})$ is a line that starts from the origin and passes through $z_{e}$. Therefore, we adopt mean squared error as loss to serve the same objective while constraining the norm of vectors in codebook $\bm{e}$:
\begin{equation}
    \label{eq:vq_loss}
    \mathcal{L}_{\mathrm{vq}} = \| z_{q} - \mathrm{sg}[z_{e}] \|^{2} + \| \mathrm{sg}[z_{q}] - z_{e} \|^{2},
\end{equation}
where the first term `pushes' $z_{q}$ to $z_{e}$, and the second term ensures the distance between the vector pairs (i.e., $z_{e}$ and $z_{q}$) does not further grow. Here, $\mathrm{sg}$ is the stop-gradient operation, which performs forward passing without partial derivatives. The assumption of our VQ module is that item features of similar characteristics and transition patterns can be learnt and aligned regardless of domain. Therefore, by training on cross-domain data, VQ learns to map item embeddings to a well-aligned feature space and alleviate the gap for transfer.

\begin{figure*}[t]
    \centering
    \includegraphics[trim=0.5cm 3.3cm 0.4cm 2.9cm, clip, width=0.75\linewidth]{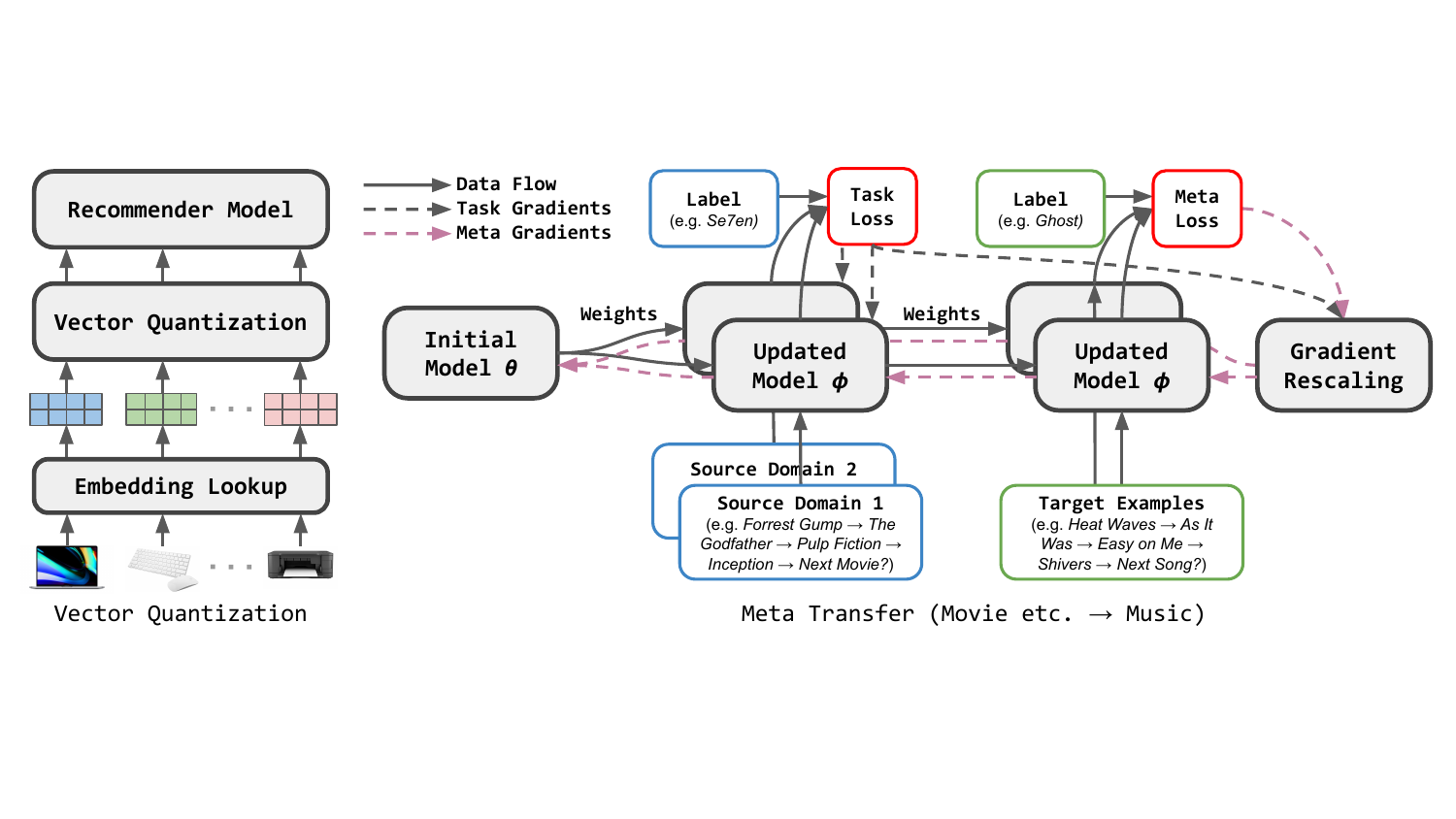}
    \caption{The proposed \ours. The left subfigure demonstrates how vector quantization is applied on sequential recommenders. The right subfigure illustrates meta transfer using multiple source domains and gradient rescaling.} 
    \label{fig:method}
    \vspace{-10pt}
\end{figure*}

\subsection{Meta Transfer}
\label{sec:meta_transfer}

% formulation
\subsubsection{Formulation}
Given $\bm{f}$ parameterized by $\bm{\theta}$, $\mathcal{X}^{s}$ and $\mathcal{X}^{t}$, meta transfer can be seen as a bi-level optimization problem:
\begin{equation}
    \label{eq:outer_lever}
    \min_{\substack{\bm{\theta}}} \mathbb{E}_{(\bm{x}^{t}, y^{t}) \sim \mathcal{X}^{t}} [\mathcal{L}(\bm{f}(\mathcal{A}lg(\bm{\theta}, \mathcal{X}^{s}); \bm{x}^{t}), y^{t})],
\end{equation}
where we seek to minimize $\mathcal{L}$ w.r.t. $\bm{\theta}$ over $\mathcal{X}^{t}$. For a collection of source datasets $\{ \mathcal{X}^{s}_{i} \}^{M}_{i=1}$, we can similarly write $\min_{\substack{\bm{\theta}}} \mathbb{E}_{\mathcal{X}^{s} \sim \{ \mathcal{X}^{s}_{i} \}^{M}_{i=1}, \, (\bm{x}^{t}, y^{t}) \sim \mathcal{X}^{t}} [\mathcal{L}(\bm{f}(\mathcal{A}lg(\bm{\theta}, \mathcal{X}^{s}); \bm{x}^{t}), y^{t})]$. This is also called outer-level optimization. Nevertheless, the original parameter set $\bm{\theta}$ is not directly used to compute the outer-level loss $\mathcal{L}$. Instead, we first optimize $\bm{\theta}$ upon source data $\mathcal{X}^{s}$ with $\mathcal{A}lg$ (e.g., gradient descent) to obtain the task-specific parameter set $\bm{\phi}$ (i.e., $\bm{\phi} = \mathcal{A}lg(\bm{\theta}, \mathcal{X}^{s})$), which is known as inner-level optimization. After that, outer-level optimization is performed to compute meta gradients w.r.t. $\bm{\theta}$. 

% inner and outer optimization
\subsubsection{Optimization}
In inner-level optimization (i.e., source task), we compute $\bm{\phi}$ upon sampled data from $\mathcal{X}^{s}$ via $\mathcal{A}lg$. $\mathcal{A}lg$ refers to some gradient descent-based optimization algorithm and is formulated as:
\begin{equation}
    \label{eq:inner_lever}
    \bm{\phi} = \mathcal{A}lg(\bm{\theta}, \mathcal{X}^{s}) = \arg\min_{\substack{\bm{\theta}}} \mathbb{E}_{(\bm{x}^{s}, y^{s}) \sim \mathcal{X}^{s}} [\mathcal{L}(\bm{f}(\bm{\theta}; \bm{x}^{s}), y^{s})].
\end{equation}
In our experiments, we sample from $\mathcal{X}^{s}$ and perform multiple steps of gradient descent with learning rate $\alpha$ to approximate \Cref{eq:inner_lever}. For each source domain, inner-level optimization only requires first-order derivatives. However, to optimize the outer-level problem, we differentiate through $\mathcal{A}lg$ (i.e., $\bm{\phi}$) back to $\bm{\theta}$, which requires second-order derivatives:
\begin{equation}
  \begin{aligned}
    \label{eq:analysis}
    \frac{d \mathcal{L}(\bm{f}(\mathcal{A}lg(\bm{\theta}, \mathcal{X}^{s}); \bm{x}^{t}), y^{t})}{d \bm{\theta}} &= \\
    \frac{d \mathcal{A}lg(\bm{\theta}, \mathcal{X}^{s})}{d\bm{\theta}} \nabla_{\bm{\phi}} &\mathcal{L}(\bm{f}(\mathcal{A}lg(\bm{\theta}, \mathcal{X}^{s}); \bm{x}^{t}), y^{t}),
  \end{aligned}
\end{equation}
recall that $\mathcal{A}lg(\bm{\theta}, \mathcal{X}^{s})$ computes $\bm{\phi}$, and thus $\frac{d\mathcal{A}lg(\bm{\theta}, \mathcal{X}^{s})}{d\bm{\theta}}$ is equivalent to $\frac{d \bm{\phi}}{d \bm{\theta}}$. The right-hand side $\nabla_{\bm{\phi}} \mathcal{L}(\bm{f}(\mathcal{A}lg(\bm{\theta}, \mathcal{X}^{s}); \bm{x}^{t}), y^{t})$ refers to first-order gradients by computing the meta loss over the sampled $(\bm{x}^{t}, y^{t})$. In this term, we consider the derivatives of the meta loss w.r.t. $\bm{\phi}$ (i.e., $\mathcal{L} \rightarrow \bm{\phi}$). However, $\frac{d\mathcal{A}lg(\bm{\theta}, \mathcal{X}^{s})}{d\bm{\theta}}$ is non-trivial as it requires second-order derivatives (i.e., Hessian matrix) to track parameter-to-parameter changes from $\bm{\phi}$ through $\mathcal{A}lg$ to the original $\bm{\theta}$. In our implementation, we differentiate the meta loss w.r.t. $\bm{\theta}$ by retaining the computational graph~\cite{finn2017model}.

% grad rescaling
\subsubsection{Gradient Rescaling}
While optimizing \Cref{eq:outer_lever} improves the target domain performance, it does so by uniformly learning without accounting for domain similarity. Therefore, we introduce a gradient rescaling algorithm that adaptively updates the parameter set $\bm{\theta}$. Specifically for the $i$-th source task, the original parameters $\bm{\theta}$ are updated with first-order derivatives as we sample from $\mathcal{X}^{s}$. As we update $\bm{\theta}$ multiple times to obtain $\bm{\phi}_{i}$, we denote the gradients of the $i$-th source task with $\bm{\phi}_{i} - \bm{\theta}$ for simplicity. Subsequently, the meta loss is computed using $\bm{\phi}_{i}$ and examples sampled from $\mathcal{X}^{t}$ (as in \Cref{eq:outer_lever}). Simplified from \Cref{eq:analysis}, we use $\frac{d\bm{\phi}_{i}}{d\bm{\theta}} \nabla_{\bm{\phi}_{i}}\mathcal{L}$ to denote the meta gradients. We compute the similarity score $s_{i}$ for the $i$-th pair of source and meta tasks with:
\begin{equation}
    \label{eq:similarity}
    s_{i} = \mathrm{Sim}(\frac{d\bm{\phi}_{i}}{d\bm{\theta}} \nabla_{\bm{\phi}_{i}}\mathcal{L}, \; \bm{\phi}_{i} - \bm{\theta}).
\end{equation}
In each training iteration, we sample $n$ (default to 3) pairs of source and target tasks and compute their similarity scores. The scores $[s_{1}, s_{2} , ..., s_{n}]$ are then transformed into a probability distribution via softmax function: $\bm{s} = \mathrm{softmax}([s_{1} / \tau, s_{2} / \tau, ..., s_{n} / \tau])$, in which we introduce temperature $\tau$ to be selected empirically. Finally, we update the parameter set $\bm{\theta}$ with rescaled gradients and learning rate $\beta$:
\begin{equation}
    \label{eq:update}
    \bm{\theta} - \beta \sum_{i}^{n} s_{i} \cdot \frac{d\bm{\phi}_{i}}{d\bm{\theta}} \nabla_{\bm{\phi}_{i}}\mathcal{L}.
\end{equation}
In our implementation, the similarity scores are computed in a \emph{per-layer} fashion. That is, we compute the scores and rescale the meta gradients individually for each of the components in the recommender model. This approach is designed to facilitate fine-grained gradient rescaling and knowledge transfer, allowing for more precise and effective adaptation.

% summarization
\subsection{Overall Framework}
Provided with the vector quantization and meta transfer modules, we illustrate the overall framework of \ours in \Cref{fig:method}. The meta objective is combined by the original recommendation loss $\mathcal{L}$ and the VQ loss $\mathcal{L}_{\mathrm{vq}}$. We use LRURec~\cite{yue2023linear} as backbone, thus the recommendation loss is cross entropy and the resulting overall loss $\mathcal{L}_{\mathrm{overall}}$ is:
\begin{equation}
\label{eq:loss}
    \mathcal{L}_{\mathrm{overall}} = \mathcal{L} + \mathcal{L}_{\mathrm{vq}},
\end{equation}
where $\mathcal{L}$ depends on the deployed recommender architecture (e.g., ranking or cross entropy loss). In each training iteration, we perform inner-level updates and compute the outer-level loss for the $n$ pairs of source and meta tasks respectively. Then, the meta gradients are rescaled and applied to $\bm{\theta}$ by computing and normalizing the similarity values.

\begin{table*}[t]
\small
\centering
% \resizebox{\textwidth}{!}{
\begin{tabular}{@{}llccccccccc@{}}
\toprule
\textbf{Dataset}                      & \textbf{Metric} & \textbf{GRU4Rec} & \textbf{NARM} & \textbf{SASRec} & \textbf{BERT4Rec} & \textbf{FMLP-Rec} & \textbf{LRURec} & \textbf{\ours}   & \textbf{Improv.} \\ \midrule
\multirow{3}{*}{\textbf{Scientific}}  & NDCG@10         & 0.0826           & 0.0843        & 0.0797          & 0.0790            & \ul{0.0995}      & 0.0979           & \textbf{0.1079} & 8.44\%           \\
                                      & Recall@10       & 0.1055           & 0.1000        & 0.1305          & 0.1061            & \ul{0.1424}      & 0.1379           & \textbf{0.1433} & 0.63\%           \\
                                      & MRR             & 0.0702           & 0.0833        & 0.0696          & 0.0759            & \ul{0.0914}      & 0.0907           & \textbf{0.1026} & 12.25\%          \\ \midrule
\multirow{3}{*}{\textbf{Instruments}} & NDCG@10         & 0.0633           & 0.0800        & 0.0634          & 0.0707            & 0.0819            & \ul{0.0853}     & \textbf{0.0903} & 5.86\%           \\
                                      & Recall@10       & 0.0969           & 0.1014        & 0.0995          & 0.0972            & 0.1092            & \ul{0.1142}     & \textbf{0.1208} & 5.78\%           \\
                                      & MRR             & 0.0707           & 0.0783        & 0.0577          & 0.0677            & 0.0789            & \ul{0.0820}     & \textbf{0.0864} & 5.37\%           \\ \midrule
\multirow{3}{*}{\textbf{Arts}}        & NDCG@10         & 0.1075           & 0.1091        & 0.0848          & 0.0942            & \ul{0.1192}      & 0.1107           & \textbf{0.1234} & 3.52\%           \\
                                      & Recall@10       & 0.1317           & 0.1315        & 0.1342          & 0.1236            & \ul{0.1543}      & 0.1471           & \textbf{0.1551} & 0.52\%           \\
                                      & MRR             & 0.1041           & 0.1060        & 0.0742          & 0.0899            & \ul{0.1136}      & 0.1045           & \textbf{0.1181} & 3.96\%           \\ \midrule
\multirow{3}{*}{\textbf{Office}}      & NDCG@10         & 0.0761           & 0.1012        & 0.0832          & 0.0972            & 0.0986            & \ul{0.1085}     & \textbf{0.1170} & 7.83\%           \\
                                      & Recall@10       & 0.1053           & 0.1203        & 0.1196          & 0.1205            & 0.1204            & \ul{0.1322}     & \textbf{0.1412} & 6.81\%           \\
                                      & MRR             & 0.0731           & 0.0984        & 0.0751          & 0.0932            & 0.0949            & \ul{0.1046}     & \textbf{0.1129} & 7.93\%           \\ \midrule
\multirow{3}{*}{\textbf{Games}}       & NDCG@10         & 0.0586           & 0.0638        & 0.0547          & 0.0628            & 0.0623            & \ul{0.0706}     & \textbf{0.0755} & 6.94\%           \\
                                      & Recall@10       & 0.0988           & 0.0977        & 0.0953          & 0.1029            & 0.0967            & \ul{0.1102}     & \textbf{0.1198} & 8.71\%           \\
                                      & MRR             & 0.0539           & 0.0609        & 0.0505          & 0.0585            & 0.0595            & \ul{0.0669}     & \textbf{0.0701} & 4.78\%           \\ \midrule
\multirow{3}{*}{\textbf{Pet}}         & NDCG@10         & 0.0648           & 0.0876        & 0.0569          & 0.0602            & 0.0829            & \ul{0.0932}     & \textbf{0.0956} & 2.58\%           \\
                                      & Recall@10       & 0.0781           & 0.1014        & 0.0881          & 0.0765            & 0.1002            & \ul{0.1108}     & \textbf{0.1136} & 2.53\%           \\
                                      & MRR             & 0.0632           & 0.0866        & 0.0507          & 0.0585            & 0.0810            & \ul{0.0913}     & \textbf{0.0932} & 2.08\%           \\ \bottomrule
\end{tabular}
% }
\caption{Evaluation results of \ours compared to ID-based baselines in cross-domain sequential recommendation. The metrics are NDCG@10, Recall@10 and MRR, with best results marked in bold and second best results underlined.}
\label{tab:id-results}
\vspace{-10pt}
\end{table*}
\section{Experiments}
\label{sec:experiment}

\subsubsection{Dataset}
We select source and target domains datasets following~\cite{ni2019justifying, hou2022learning, li2023text}. In particular, we adopt \emph{Automotive}, \emph{Cell Phones and Accessories}, \emph{Clothing Shoes and Jewelry}, \emph{Electronics}, \emph{Grocery and Gourmet Food}, \emph{Home and Kitchen}, \emph{Movies and TV} and \emph{CDs and Vinyl} as our source datasets (i.e., Source). For target datasets, we select \emph{Industrial and Scientific} (Scientific), \emph{Musical Instruments} (Instruments), \emph{Arts, Crafts and Sewing} (Arts), \emph{Office Products} (Office), \emph{Video Games} (Games) and \emph{Pet Supplies} (Pet). For preprocessing, we follow previous works~\cite{hou2022towards, yue2022defending, hou2022learning, li2023text} by performing k-core filtering.

\subsubsection{Evaluation} Following~\cite{yue2022defending, hou2022learning, li2023text}, we adopt the leave-one-out approach, which uses the last two items in each sequence for validation and test. We adopt normalized discounted cumulative gain (NDCG@$k$), recall (Recall@$k$) and mean reciprocal rank (MRR) with $k = 10$ as metrics. We save the model with best validation NDCG@10 scores for evaluation on the test set. We compute the metric values by ranking the ground-truth item against all items in the target dataset. For baselines, we select GRU4Rec~\cite{hidasi2015session}, NARM~\cite{li2017neural}, SASRec~\cite{kang2018self}, BERT4Rec~\cite{sun2019bert4rec}, FMLP-Rec~\cite{zhou2022filter} and LRURec~\cite{yue2023linear}.

\subsection{RQ1: How does \ours perform in cross-domain sequential recommendation?}
\label{sec:rq1}

% comparison to id-based methods
We first evaluate the cross-domain recommendation performance of \ours with other \textit{ID-based} baseline methods. The evaluation results for each target dataset are reported in \Cref{tab:id-results}. Overall, the baselines are consistently outperformed by \ours, confirming the effectiveness of the proposed \ours in cross-domain recommendation. Specifically, we observe:
(1)~\ours performs the best across all scenarios, successfully improving target domain performance without requiring auxiliary information. On average, \ours achieves $6.36\%$ improvements on NDCG@10 compared to the best-performing baseline. 
(2)~\ours shows significant improvements on Office and Games ($7.39\%$ on NDCG@10), while achieving moderate gains on Arts and Pet ($3.05\%$ on NDCG@10). These results suggest that \ours may perform differently across domains.
(3)~In contrast to recall scores, \ours demonstrates a better ranking performance. For instance on the Scientific dataset, the performance on NDCG@10 increases by $8.44\%$, while the relative improvement on Recall@10 is lower at $0.63\%$.
Overall, the results in \Cref{tab:id-results} show a significantly improved transfer performance by \ours, suggesting the efficacy of the proposed method.

\subsection{RQ2: What contributes to the performance of \ours?}

\begin{table*}[t]
\small
\centering
\begin{tabular}{@{}llcccccc@{}}
\toprule
\textbf{Method}                                        & \textbf{Metric} & \textbf{Scientific} & \textbf{Instruments} & \; \textbf{Arts} \; & \; \textbf{Office} \; & \; \textbf{Games} \; & \; \textbf{Pet} \;\; \\ \midrule
\multirow{3}{*}{\textbf{\ours}}                        & NDCG@10         & \textbf{0.1079}     & \textbf{0.0903}      & \textbf{0.1234} & \textbf{0.1170} & \textbf{0.0755} & \textbf{0.0956} \\
                                                       & Recall@10       & \textbf{0.1433}     & \textbf{0.1208}      & \textbf{0.1551} & \textbf{0.1412} & \textbf{0.1198} & \textbf{0.1136} \\
                                                       & MRR             & \textbf{0.1026}     & \textbf{0.0864}      & \textbf{0.1181} & \textbf{0.1129} & \textbf{0.0701} & \textbf{0.0932} \\ \midrule
\multirow{3}{*}{\textbf{\ours w/o multi-head VQ}}      & NDCG@10         & 0.1050              & 0.0876               & 0.1177          & 0.1109          & 0.0703          & 0.0923          \\
                                                      & Recall@10       & 0.1406              & 0.1165               & 0.1476          & 0.1331          & 0.1087          & 0.1093          \\
                                                      & MRR             & 0.0992              & 0.0839               & 0.1131          & 0.1072          & 0.0662          & 0.0904          \\ \midrule
\multirow{3}{*}{\textbf{\ours w/o VQ}}                 & NDCG@10         & \ul{0.1076}        & \ul{0.0895}         & 0.1175          & \ul{0.1146}    & \ul{0.0739}    & 0.0923          \\
                                                      & Recall@10       & \ul{0.1437}        & \ul{0.1183}         & 0.1449          & \ul{0.1379}    & \ul{0.1175}    & 0.1097          \\
                                                      & MRR             & \ul{0.1018}        & \ul{0.0861}         & 0.1135          & \ul{0.1105}    & 0.0687          & 0.0903          \\ \midrule
\multirow{3}{*}{\textbf{\ours w/o gradient rescaling}} & NDCG@10         & 0.1070              & 0.0886               & \ul{0.1214}    & 0.1134          & 0.0738          & \ul{0.0926}    \\
                                                      & Recall@10       & 0.1409              & 0.1178               & \ul{0.1507}    & 0.1367          & 0.1167          & \ul{0.1102}    \\
                                                      & MRR             & 0.1016              & 0.0853               & \ul{0.1169}    & 0.1095          & \ul{0.0688}    & \ul{0.0906}    \\ \midrule
\multirow{3}{*}{\textbf{\ours w/o meta transfer}}      & NDCG@10         & 0.1019              & 0.0721               & 0.1089          & 0.0952          & 0.0676          & 0.0868          \\
                                                      & Recall@10       & 0.1345              & 0.0947               & 0.1354          & 0.1121          & 0.1040          & 0.1012          \\
                                                      & MRR             & 0.0973              & 0.0696               & 0.1048          & 0.0926          & 0.0641          & 0.0856          \\ \bottomrule
\end{tabular}
\caption{Ablation results of \ours, with best results marked in bold and second best results underlined.}
\label{tab:ablation}
\vspace{-10pt}
\end{table*}

% intro to ablation
In this research question, we evaluate the effectiveness of \ours by ablating the proposed method. Specifically, we study variants of \ours to assess the effectiveness of individual modules. We report the performance of \ours and its variants in \Cref{tab:ablation}, including: (1)~\ours without multi-head VQ; (2)~\ours without VQ; (3)~\ours without gradient rescaling; and (4)~We additionally substitute meta transfer with joint training (i.e., \ours w/o meta transfer).
We observe the following: 
(1)~the proposed multi-head VQ performs well in aligning item features. In contrast, substituting the multi-head approach or removing VQ causes consistent performance drops, suggesting the effectiveness of employing multi-head VQ while sharing weights with target domain embeddings.
(2)~Removing gradient rescaling or meta transfer also leads to consistent performance deterioration across metrics. On average, $2.19\%$ NDCG@10 improvements can be attributed to the proposed gradient scaling, whereas removing meta transfer causes $14.86\%$ NDCG@10 drops.
In summary, the ablation results confirm the effectiveness of the parameter-efficient VQ and meta transfer mechanisms in \ours, consistently enhancing recommendation performance in cross-domain transfer scenarios.
\section{Conclusion}

In this work, we investigate an \emph{ID-only}, \emph{non-overlapping} and \emph{multi-source} setting for universal transfer learning on sequential recommenders. In particular, we design vector quantized meta transfer for sequential recommenders (\ours). The VQ module is designed to map domain-specific item embeddings into a shared feature space. Moreover, the proposed meta transfer adaptively learns from the source domains to guide the transfer of source knowledge to the target domain. As such, \ours maximizes the transfer learning performance via generalizable representations and exploitation of the source domains. We demonstrate the effectiveness of \ours on multiple datasets, where \ours consistently outperforms state-of-the-art baseline methods by a considerable margin.

\bibliographystyle{IEEEtran}
\bibliography{reference}

\end{document}